\documentclass[pra,twocolumn,showpacs,superscriptaddress]{revtex4}
\usepackage{graphicx}
\usepackage{dcolumn}
\usepackage{amsmath}
\usepackage{amssymb}
\usepackage{amsmath,verbatim}
\usepackage{bm}
\def\rv{{\bf r}}

\def\nn{\nonumber}
\begin{document}
\title{Performance of a non-empirical meta-GGA density functional
for excitation energies} 
\author{Jianmin Tao}
\email{jtao@lanl.gov}
\affiliation{Theoretical Division and Center for Nonlinear Studies, 
Los Alamos National Laboratory, Los Alamos, New Mexico 87545}
\author{Sergei Tretiak}
\affiliation{Theoretical Division and Center for Nonlinear Studies, 
Los Alamos National Laboratory, Los Alamos, New Mexico 87545} 
\affiliation{Center for Integrated
Nanotechnology, Los Alamos National Laboratory, Los Alamos, New
Maxico 87545}
\author{Jian-Xin Zhu}
\email{jxzhu@lanl.gov}
\homepage{http://theory.lanl.gov}
\affiliation{Theoretical Division and Center for Nonlinear Studies, 
Los Alamos National Laboratory, Los Alamos, New Mexico 87545}

\date{\today}
\begin{abstract}
It is known that the adiabatic approximation in time-dependent density 
functional theory usually provides a good description of low-lying 
excitations of molecules. In the present work, the capability of the 
adiabatic nonempirical meta-generalized gradient approximation (meta-GGA) 
of  Tao, Perdew, Staroverov, and Scuseria (TPSS)
to describe atomic and molecular excitations is tested. The 
adiabatic (one-parameter) hybrid version of the TPSS meta-GGA and the adiabatic
GGA of Perdew, Burke, and Ernzerhof (PBE) are also included in the test. 
The results are compared to experiments and to two well-established 
hybrid functionals PBE0 and B3LYP.
Calculations show that both adiabatic TPSS and TPSSh functionals produce   
excitation energies in fairly good agreement with experiments, and improve
upon the adiabatic local spin density approximation and, in particular, the
adiabatic PBE GGA. 
This further confirms that TPSS is indeed a reliable 
nonhybrid universal functional which can serve as the starting point from 
which higher-level approximations can be constructed.
The systematic underestimate of the low-lying vertical excitation energies
of molecules with time-dependent density functionals within the
adiabatic approximation suggests that further improvement can be made
with nonadiabatic corrections.

\pacs{
71.15.Mb, 31.15.Ew, 71.45.Gm}
\end{abstract}
\maketitle

\section{Introduction}
Density functional theory (DFT)~\cite{ks65,Parr-Yang,Perdew-Kurth} 
is a mainstream electronic structure theory of many-electron systems, 
which has achieved a high-level sophitication. While a ladder of 
highly accurate exchange-correlation (xc) 
functionals~\cite{pbe96,tpss,prtssc} 
have been constructed, they are essentially not suitable for
the description of time-dependent processes and excited states, 
because these density functionals are only protected by the 
ground-state Hohenberg-Kohn variational principle. This limitation 
has been overcome by the most important extension of DFT-- time-dependent 
density functional theory (TDDFT)~\cite{rg}. 

In recent years TDDFT has 
been rapidly grown into a popular method for the investigation of dynamical 
properties of many-electron systems~\cite{gvbook,grossbook,sergei1,sergei2}. 
It follows the Kohn-Sham strategy
and maps the complicated problem of interacting electrons in 
a time-dependent external potential $v(\rv,t)$ to a simpler 
one of noninteracting  electrons moving in a self-consistent 
time-dependent effective potential $v_{\rm s}(\rv,t) = 
v(\rv,t) + u_{\rm H}(\rv,t) + v_{\rm xc}(\rv,t)$, which generates
the density $n(\rv,t)$ of the interacting system. Here 
 $u_{\rm H}(\rv,t)$ is the Hartree potential given by
$u_{\rm H}(\rv,t) = \int d^3r' n(\rv',t)/|\rv - \rv'|$, and
$v_{\rm xc}(\rv,t)$ is the dynamical XC potential 
defined by $v_{\rm xc}(\rv, t) \equiv 
\delta A_{\rm xc}[n]/\delta n(\rv,t)$,
with $A_{\rm xc}[n]$ 
being the time-dependent XC functional or XC action, 
the analogue of the static functional $E_{\rm xc}[n_0]$.
In the linear response, 
$v_{\rm s}(\rv,t) = v_{\rm s,0}(\rv) + v_{\rm s,1}(\rv,t)$,
where the effective ground-state potential $v_{\rm s,0}(\rv)$
can be written as the sum of the external
potential, the Hartree potential, and the XC potential of the
ground state, i.e.,
$v_{\rm s,0}(\rv) = v_0(\rv) + u_{\rm H,0}(\rv) + v_{\rm xc,0}(\rv)$,
while the effective perturbation $v_{\rm s,1}(\rv,t)$ is the sum of the 
external (or physical) perturbation, the induced Hartree and XC potentials,
$v_{\rm s,1}(\rv,t) = v_1(\rv,t) + u_{\rm H,1}(\rv,t) + v_{\rm xc,1}(\rv,t)$.

In TDDFT, the physical excitations
can be calculated from the 
linear response theory~\cite{casida95,pgg1999} 
through the electron 
density-density response function~\cite{gvbook} 
$\chi(\rv,\rv',t,t')$, in which the 
only unknown part is the XC kernel defined by
\begin{eqnarray}\label{kernel}
f_{\rm xc}(\rv,\rv',t,t') \equiv \frac{\delta v_{\rm xc}([n];\rv, t)}
{\delta n(\rv' ,t')}.
\end{eqnarray}
The key idea is that the exact linear density response of an 
interacting system to the external perturbation can be 
written as the linear density response of a noninteracting system
to the effective perturbation, i.e,
\begin{eqnarray}\label{density_response}
n_1(\rv,\omega) &=& \int d\rv~ \chi(\rv,\rv',\omega)
v_{\rm 1}(\rv',\omega) \nn \\
&=& \int d\rv~ \chi_{\rm s}(\rv,\rv',\omega)
v_{\rm s,1}(\rv',\omega)~, 
\end{eqnarray} 
where
\begin{eqnarray}\label{effective1}
v_{\rm s,1}(\rv,t) &=& v_1(\rv,t) + 
\int d\rv' \frac{n_1(\rv',t)}{|\rv - \rv'|} \nn \\
&+&
\int dt' \int d\rv' f_{\rm xc}(\rv,\rv',t,t') n_1(\rv',t),
\end{eqnarray}
and $\chi_{\rm s}(\rv,\rv',\omega)$ is the Kohn-Sham response function
evaluated with the Kohn-Sham ground-state orbitals. For spin-unpolarized
systems, we have
\begin{eqnarray}
\chi_{\rm s}(\rv,\rv'; \omega) = 
2 \sum_{j,k}(n_k - n_j)\frac{\phi_k^*(\rv)\phi_j(\rv)
\phi_j^*(\rv')\phi_k(\rv')}{\omega - \omega_{jk} + i\delta}, 
\end{eqnarray}
where $n_k$ are the orbital electron occupation numbers. 
By substituting the effective perturbation $v_{\rm s,1}(\rv,t)$ into 
Eq.~(\ref{density_response}) with the observation that
the poles $\omega_{jk}$ of the  Kohn-Sham response 
function are generally different from those of the interacting system,
one arrives at an equation~\cite{pgg1999}, from which excitation 
energies of the interacting system can be calculated as an 
eigenvalue problem.

As in the static DFT, the dynamical XC potential must be approximated
in practice. The simplest construction is the adiabatic 
approximation ~\cite{zs}, which makes use of the ground-state XC potential 
but replaces the ground-state density $n_0(\rv)$ with the 
instantaneous density $n(\rv,t)$, namely,
\begin{eqnarray}\label{adiabatic}
v_{\rm xc}^{ad}([n];\rv,t) =
\frac{\delta E_{\rm xc}[n_0]}{\delta n_0(\rv)}
\Big |_{n_0(\rv) = n(\rv,t)}~.
\end{eqnarray}
Within the adiabatic approximation the XC kernel can be calculated from
\begin{eqnarray}\label{adkernel}
f_{\rm xc}(\rv,\rv',t,t') \equiv \frac{\delta v_{\rm xc}([n_0];\rv)}
{\delta n(\rv')}\delta(t - t'),
\end{eqnarray}
which is local in time, while it is not necessarily local in space.
This approximation, which completely neglects the vector 
potential~\cite{vk96,tokatly05,tv06} 
and thus retardation and dissipation effects~\cite{tvt071,dv06,ullrich06}, 
has been widely used to 
calculate the single-particle excitation 
energies~\cite{Ahlrichs96,handy98,gus98,gus99,sggb00,gorling03,jbp,sergei3,yang07},
although it fails for multi-particle excitations~\cite{th,tahra}
or charge transfer states\cite{dreuw,tozer,neepa}, 
due to the disregard of the frequency dependence.
The detail of the TDDFT linear response theory for the calculation 
of the excitation energies within the adiabatic approximation has been 
documented in the literature~\cite{casida95,Ahlrichs96,pgg1999}.

Although in general a nonadiabatic correction to the adiabatic
approximation is needed even in the low-frequency limit, 
it has been shown~{\cite{pgg1999,gksg98}
that (at least for small systems) the largest source of error in the 
prediction of low-lying excitation energies arises from the 
approximation to the static XC potential. This justifies the 
adiabatic approximation for the description of low-lying excitations 
of atoms and molecules. The low-lying excited states in the visible
and near-UV region are the most interesting ones. For example,
photodissociation often proceeds on the lowest excited potential 
energy surface, and the photoemmision wavelength of materials is 
controlled by the lowest electronic excitations. 
A quantitative description of electronic excitated
states of molecules is important in spectroscopy, photochemistry,
and the design of optical materials (e.g., design of dyes).  
Therefore, assessment of the ability 
of time-dependent density functionals in 
describing electronic excitations is of general interest.

In the present work,
the capability of the adiabatic nonempirical meta-generalized
gradient approximation (meta-GGA) of Tao, Perdew, Staroverov, and 
Scuseria (TPSS)~\cite{tpss,ptss} and its adiabatic one-parameter hybrid 
version~\cite{sstp1} to describe low-lying excitations is  
tested for eleven atoms with $Z \le 36$ and 
prototype small molecules CO, N$_2$, H$_2$O, CH$_2$O (formaldehyde), 
(CH$_3$)$_2$CO (acetone), C$_2$H$_4$ (ethylene), C$_6$H$_6$ (benzene), 
and C$_5$H$_5$N (pyridine). 
The reason for choosing these atoms and molecules as
our test set is that the high-quality experimental results of these 
systems are available. The TPSS hybrid functional (TPSSh)~\cite{sstp1}
\begin{eqnarray}\label{global}
E_{\rm xc}^{\rm TPSSh} = aE_{\rm x}^{\rm exact} + (1 -a)E_{\rm x}^{\rm TPSS} 
+ E_{\rm c}^{\rm TPSS},
\end{eqnarray}
contains a global empirical parameter -- the exact-exchange mixing 
coefficient ``$a = 0.1$'', which is determined by a fit to 
experimental atomization energies of molecules. With the 
introduction of exact exchange, TPSSh functional, however, does not 
satisfy any universal constraints beyond those satisfied by the 
TPSS meta-GGA, but it improves the description of the
asymptotic behavior of the TPSS potential. This improvement turns out
to be helpful in most cases~\cite{tpss}. 

Since the TPSS meta-GGA is constructed from the 
GGA of Perdew, Burke, and Ernzerhof (PBE)~\cite{pbe96}, and PBE GGA is  
constructed from the local spin density approximation (LSDA), we also
include the adiabatic LSDA and the adiabatic PBE GGA in the present 
test. Then we compare the results to both experiments and two 
well-established hybrid functionals: PBE0 and B3LYP, former of which 
is a one-parameter hybrid functional~\cite{gus99} based 
on the PBE GGA with the choice~\cite{perdew96} of 1/4 as its exact-exchange 
mixing coefficient, and the latter is a three-parameter empirical 
hybrid functional~\cite{B93} with 1/5 exact exchange mixed in its
exchange component.           
Numerical results show that both adiabatic 
TPSS and TPSSh yields low-lying excitation energies of atoms and 
molecules with fairly good accuracy. In particular, we find that the 
adiabatic TPSS meta-GGA improves upon the PBE GGA and even the adiabatic 
LSDA, the latter of which usually works well for molecular excitations. 
The adiabatic TPSSh consistently yields further improvement and can
achieve the comparable accuarcy of the most popular hybrid functional
B3LYP.   
%%%%%%%%%%%%%%%%%%%%%%%%%%%%%%%
\begin{table*}
\caption{ Two lowest-lying singlet excitation energies (in eV) of atoms
calculated using six functionals with the basis set 6-311++G(3df,3pd).
The mean error (m.e.) (with the sign convention that error =  
theory - experiment) and the mean absolute error (m.a.e.) 
are also shown. The mean experimental value of these atoms is 8.06 eV.
(1 hartree = 27.21 eV).
}
\begin{tabular}{llccccccc}
\hline \hline 
Atom &  Transition & LSD & PBE & TPSS & TPSSh &PBE0&B3LYP& Expt$^a$
\\ \hline
He& $1s \rightarrow 2s$ & 19.59 &19.73 &20.27 &20.58 &20.62 &20.50 &20.62  \\
  & $1s \rightarrow 2s$ & 22.99 &23.41 &24.04 &24.23 &24.05 &23.95 &21.22  \\
Li& $2s \rightarrow 2p$ & 1.98  &1.98  &1.99  &1.97  &1.95  &1.98 & 1.85\\
  & $2s \rightarrow 3s$ & 3.12  &3.09  &3.09  &3.13  &3.23 &3.16 & 3.37 \\
Be& $2s \rightarrow 2p$ & 4.84  &4.91  &5.06  &5.05  &4.94 &4.88 & 5.28\\
  & $2s \rightarrow 3s$ & 6.11  &6.12  &6.29  &6.35  &6.32 &6.21 & 6.78\\
Ne& $2p \rightarrow 3s$ &17.45  &17.21 &17.55 &17.94 &18.27 &17.88 & 16.62 \\
  & $2p \rightarrow 3p$ &19.82  &19.46 &19.74 &20.16 &20.59 &20.11 & 18.38 \\
Na& $3s \rightarrow 3p$ & 2.25  &2.12  &2.02  &2.02  &2.08 &2.23 & 2.10\\
  & $3s \rightarrow 4s$ & 3.05  &2.91  &2.87  &2.90  &3.02 &3.02 & 3.19\\
Mg& $3s \rightarrow 3p$ & 4.24  &4.18  &4.18  &4.19  &4.20 &4.23 & 4.35\\
  & $3s \rightarrow 4s$ & 5.02  &4.93  &5.01  &5.06  &5.08 &5.00 & 5.39\\
Ar& $3p \rightarrow 4s$ &11.32  &11.27 &11.59 &11.81 &11.90 &11.56 &11.55 \\
  & $3p \rightarrow 4p$ &12.68  &12.50 &12.74 &13.00 &13.22 &12.89 &12.91 \\
K & $4s \rightarrow 4p$ & 1.70  &1.50  &1.36  &1.36  &1.45 &1.64 & 1.61\\
  & $4s \rightarrow 5s$ & 2.52  &2.35  &2.28  &2.30  &2.42 &2.43 & 2.61\\
Ca& $4s \rightarrow 3d$ & 1.88  &1.88  &1.87  &2.02  &2.24 &2.16 & 2.71\\
  & $4s \rightarrow 4p$ & 3.09  &2.98  &2.90  &2.90  &2.96 &3.03 & 2.93\\
Zn& $4s \rightarrow 4p$ & 5.80  &5.67  &5.59  &5.52  &5.51 &5.65 & 5.80 \\
  & $2s \rightarrow 5s$ & 6.38  &6.12  &6.10  &6.12  &6.20 &6.22 & 6.92 \\
Kr& $4p \rightarrow 5s$ &9.52   &9.43  &9.72  &9.92  &10.01 &9.69 & 9.92 \\
  & $4p \rightarrow 5p$ &10.84  &10.64 &10.85 &11.10 &11.30 &10.98 &11.30 \\
m.e. & &-0.06 &-0.14 &-0.02 &0.12&0.19 &0.09 &... \\
m.a.e. & &0.47 &0.51 &0.49 &0.50 &0.50 &0.47 &...\\
\hline \hline
\end{tabular}

\footnotesize{$^a$From Ref.~\cite{Moore1971}.}
\end{table*}
%%%%%%%%%%%%%%%%%%%%%%%%%%%%%  

\section{Computational method}
All calculations are performed using the GAUSSIAN 03
suite~\cite{G03}. Vertical excitation energies of molecules are 
self-consistently calculated using the self-consistent ground-state 
geometries optimized with respective density functionals and 
the same basis set as used in the geometry optimization. 
In order for the test to be reliable, a 
fairly large basis set should be employed. Since our calculations 
involve the treatment of both ground state (optimization of molecular 
geometry) and excited states (with the adiabatic approximation within TDDFT), 
for consistency we chose the same basis 6-311++G(3df,3pd) in all 
calculations. This large basis set was previously used to perform a 
comprehensive assessment~\cite{sstp1} of the TPSS functional for molecules. 
The ultrafine grid (Grid=UltraFine) in numerical integration
and the tight self-consistent field convergence criterion
(SCF=Tight) are used. 

To make our comparisions to be consistent, 
we perform our own calculations with all reference functionals, rather
than attempting to extract data from the literature. Throughout the paper, 
we calculate the mean error (m.e.) (or signed error) using the sign convention:
error = theory - experiment. The mean error tells us whether excitation  
energies are underestimated or overestimated on the average with a particular 
density functional, while the mean absolute error (m.a.e.) shows us how far
a density functional theoretical estimate is from experiment.

\section{Results and discussions}
\subsection{Atoms}

As a simple test, we present two lowest-lying singlet excitation energies of
eleven atoms with $Z \le 36$: He, Li, Be, Ne, Na, Mg, Ar, K, Ca, Zn,
and Kr. They are calculated with the adiabatic LSDA, PBE GGA, PBE0, B3LYP,
TPSS meta-GGA, and TPSSh functionals. The results are shown in Table 1.
Experimental values~\cite{Moore1971} are also listed for comparision.

Table 1 shows that the six adiabatic density functionals
tested here produce remarkably accurate excitation energies, all with
mean absolute error (m.a.e.) of 0.5 eV.
From the mean errors (m.e.), 
we can see that all nonhybrid functionals (LSD, PBE GGA, and TPSS meta-GGA)
slightly underestimate low-lying excitation energies of atoms, while
all hybrid functionals (PBE0, B3LYP, and TPSSh) yield overestimates. 
This suggests the difficulty 
of further systematic improvement from the nonadiabatic 
corrections~\cite{vk96,tv06}. This is resonant with the numerical studies
of atomic excitation energies made by Ullrich and Burke~\cite{ub04}. 

\subsection{Molecules}
Theoretical prediction or interpretation of discrete molecular 
electronic excitation spectrum is of significant importance.
Many physical and chemical properties of materials 
are directly related to electronic excitations.    
In this work, we calculate several low-lying excitation energies
of our test set, which includes three inorganic (CO, N$_2$, H$_2$O) and
five organic (CH$_2$O, (CH$_3$)$_2$CO, C$_2$H$_4$, benzene, pyridine) 
molecules. The results are reported in Tables II--IX, respectively.

The performance of the adiabatic 
LSDA~\cite{Ahlrichs96,handy98,gus98,sggb00}, 
PBE GGA~\cite{gus99}, PBE0~\cite{gus99}, and 
B3LYP~\cite{Ahlrichs96,handy98,gus98,gus99} functionals 
has been also tested with other bases and discussed elsewhere. 
In the present work, we shall focus only on the adiabatic TPSS 
and TPSSh functionals.    

%%%%%%%%%%%%%%%%%%%%%%%%%%%%%%%%%%%%%%%%%%%%%%%%%%%%%%
\begin{table}
\caption{ Low-lying excitation energies (in eV) of CO
calculated using six functionals with the basis set 6-311++G(3df,3pd).
Calculations are performed using the geometry optimized on respective
functionals with the same basis.
The mean error (m.e) and the mean absolute error (m.a.e.) are also shown.
The mean experimental value is 9.58 eV.
%1 hartree = 27.2116 eV.
}
\begin{tabular}{llccccccc}
\hline \hline
 Symmetry&LSD & PBE & TPSS & TPSSh & PBE0& B3LYP& Expt$^a$
\\ \hline
$^3\Pi$        & 5.98  &5.68 &5.75 &5.78 &5.77 &5.89 & 6.32  \\
$^3\Sigma^{+}$ & 8.45  &7.97 &7.88 &7.88 &7.96 &8.03 &8.51  \\
$^1\Pi$        & 8.19  &8.19 &8.40 &8.50 &8.49 &8.47 &8.51  \\
$^3\Delta$     & 9.21  &8.59 &8.53 &8.59 &8.70 &8.71 &9.36  \\
$^3\Sigma^{-}$ & 9.90  &9.31 &9.64 &9.92 &9.89 &9.80 &9.88  \\
$^1\Sigma^{-}$ & 9.94  &9.79 &10.05&10.15&9.89 &9.86 & 9.88  \\
$^1\Delta$     & 9.90  &9.72 &9.96 &10.01&10.29&10.26 &10.23  \\
$^3\Sigma^{+}$ &9.55   &9.72 &9.96 &10.01&10.05&9.92 &10.40  \\
$^3\Sigma^{+}$ &10.48  &10.21&10.60&10.86&10.94&10.85 &11.30  \\
$^1\Sigma^{+}$ &10.73  &10.62&10.89&11.15&11.31&11.32 &11.40  \\
  & & & & & \\
m.e. &-0.35  &-0.60 &-0.41 &-0.30 &-0.25 &-0.28 &... \\
m.a.e.  &0.36  &0.60 &0.45 &0.36 &0.27 &0.28 &... \\
\hline \hline
\end{tabular}

\footnotesize{$^a$From Ref.~\cite{nielsen80}.}
\end{table}
%%%%%%%%%%%%%%%%%%%%%%%%%%%%%%%%%%%%%%%%%%%%%%%%%%%%%%%%%%%
\begin{table}
\caption{ Low-lying excitation energies (in eV) of N$_2$
calculated using six functionals with the basis set 6-311++G(3df,3pd).
Calculations are performed using the geometry optimized on respective
functionals with the same basis.The mean experimental value is 9.38 eV.}
\begin{tabular}{llccccccc}
\hline \hline
Symmetry& LSD & PBE & TPSS & TPSSh & PBE0& B3LYP &Expt$^a$
\\ \hline
$^3\Sigma_{\rm u}^+$ & 7.96  &7.42 &7.22 &7.12 &7.14 &7.25&7.75  \\
$^3\Pi_{\rm g}$      & 7.62  &7.34 &7.43 &7.54 &7.64 &7.68&8.04  \\
$^3\Delta_{\rm u}$   & 8.90  &8.19 &8.05 &8.01 &8.06 &8.12&8.88  \\
$^1\Pi_{\rm g}$      & 9.11  &9.04 &9.23 &9.37 &9.43 &9.37&9.31  \\
$^3\Sigma_{\rm u}^-$ & 9.73  &9.58 &9.82 &9.79 &9.53 &9.47&9.67  \\
$^1\Sigma_{\rm u}^-$ & 8.73  &9.58 &9.82 &9.79 &9.53 &9.47&9.92  \\
$^1\Delta_{\rm u}$   & 10.28 &9.98 &9.95 &9.98 &10.05 &10.86&10.27  \\
$^3\Pi_{\rm u}$      &10.39  &10.37&10.65&10.79&10.79 &10.68& 11.19  \\
  & & & & & \\
m.e. &-0.29  &-0.44 &-0.36 &-0.33 &-0.36 &-0.27 &... \\
m.a.e.  &0.36 &0.44 &0.40 &0.38 &0.39 &0.43 &... \\
\hline \hline
\end{tabular}

\footnotesize{$^a$From Ref.~\cite{bk90}.}
\end{table}
%%%%%%%%%%%%%%%%%%%%%%%%%%%%%%%%%%%%%%%%%%%%%%%%%%%
\begin{table}
\caption{ Low-lying excitation energies (in eV) of H$_2$O
calculated using six functionals with the basis set 6-311++G(3df,3pd).
Calculations are performed using the geometry optimized on respective
functionals with the same basis. The mean experimental value is 8.99 eV.}
\begin{tabular}{llccccccc}
\hline \hline
Symmetry & LSD & PBE & TPSS & TPSSh & PBE0& B3LYP & Expt$^a$
\\ \hline
$^3{\rm B}_1$ & 6.30&6.06   &6.30  &6.59 &6.80&6.56&7.14 \\
$^1{\rm B}_1$ &6.60 &6.44  &6.65 &6.96   &7.24&6.96&7.49 \\
$^3{\rm A}_2$  &7.99 &7.72  &7.90 &8.24  &8.57&8.31&9.1  \\
$^1{\rm A}_2$  &8.08 &7.88  &8.05 &8.39  &8.77&8.47&9.2 \\
$^3{\rm A}_1$ &8.26 &8.10 &8.36 &8.64    &8.84&8.58& 9.35\\
$^1{\rm A}_1$  &8.67 &8.62  &8.86 &9.15  &9.43&9.10&9.73 \\
$^3{\rm B}_2$  &9.94 &9.75  &9.95 &10.26 &10.55&10.28&9.93\\
$^1{\rm B}_2$ &10.14 &10.04 &10.23 &10.57&10.93&10.59&10.0 \\
  & & & & & \\
m.e. &-0.75 &-0.92 &-0.71 &-0.39 &-0.10 &-0.39 &... \\
m.a.e.  &0.78 &0.93 &0.77 &0.62 &0.49 &0.62 &... \\
\hline \hline
\end{tabular}

\footnotesize{$^a$From Ref.~\cite{kaldor87}.}
\end{table}
%%%%%%%%%%%%%%%%%%%%%%%%%%%%%%%%%%%%%%%%%%%%%%%%%%%%%%%%%%%%%%
\begin{table}
\caption{ Low-lying excitation energies (in eV) of
formaldehyde (H$_2$CO)
calculated using six functionals with the basis set 6-311++G(3df,3pd).
Calculations are performed using the geometry optimized on respective
functionals with the same basis. The mean experimental value is 6.90 eV.}
\begin{tabular}{llccccccc}
\hline \hline
Symmetry & LSD & PBE & TPSS & TPSSh & PBE0& B3LYP & Expt$^a$
\\ \hline
$^3$A$_2$ &  3.15  &3.09  &3.26 &3.30&3.22&3.26 & 3.5  \\
$^1$A$_2$ &  3.75  &3.82  &4.06 &4.12&4.02&3.99 & 4.1  \\
$^3$A$_1$ &  6.37  &5.75  &5.57 &5.46&5.43&5.58 & 6.0  \\
$^3$B$_2$ &  5.89  &5.68  &5.95 &6.27&6.53&6.38 & 7.09  \\
$^1$B$_2$ &  5.99  &5.89  &6.11 &6.45&6.77&6.53 & 7.13  \\
$^3$B$_2$ &  7.10 &6.91 &7.17 &7.44  &7.62&7.46 & 7.92  \\
$^1$B$_2$ &  7.18 &7.07 &7.29 &7.58  &7.82&7.61 & 7.98  \\
$^3$A$_1$ &  6.86 &6.63 &6.87 &7.21  &7.50&7.35 &8.11  \\
$^1$A$_1$ & 6.95  &6.82  &7.01 &7.36 &7.72&7.47 & 8.14  \\
$^1$B$_1$ & 8.86  &8.82 &9.01 &9.15  &9.22&9.09 &9.0  \\
  & & & & & \\
m.e. &-0.69 &-0.87 &-0.69 &-0.49 &-0.31 &-0.43 &... \\
m.a.e.  &0.77 &0.87 &0.69 &0.52 &0.36 &0.44 &... \\
\hline \hline
\end{tabular}

\footnotesize{$^a$From Refs.~\cite{pople92,hfw93}.}
\end{table}
%%%%%%%%%%%%%%%%%%%%%%%%%%%%%%%%%%%%%%%%%%%%%%%%%%%
%%%%%%%%%%%%%%%%%%%%%%%%%%%%%%%%%%%%%%%%%%%%%%%%%%%%%%%%%%%%%%
\begin{table}
\caption{ Low-lying excitation energies (in eV) of
acetone ((CH$_3$)$_2$CO)
calculated using six functionals with the basis set 6-311++G(3df,3pd).
Calculations are performed using the geometry optimized on respective
functionals with the same basis. The mean experimental value is 6.17 eV.}
\begin{tabular}{llccccccc}
\hline \hline
Symmetry & LSD & PBE & TPSS & TPSSh & PBE0& B3LYP & Expt$^a$
\\ \hline
$^3{\rm A}_2$ &3.70 &3.59 &3.69 &3.73&3.81&3.81 & 4.18  \\
$^1{\rm A}_2$ &4.22 &4.21 &4.37 &4.41&4.49&4.44 & 4.43 \\
$^3{\rm A}_1$ &6.13 &5.70 &5.97 &5.96&5.60&5.70 & 5.88\\
$^3{\rm A}_2$ &6.28 &6.11 &6.27 &6.26&6.01&5.75 & 6.26  \\
$^1{\rm B}_2$ &5.09 &5.00 &5.22 &5.22&6.08&5.80 & 6.36  \\
$^1{\rm A}_2$ &6.30 &6.14 &6.30 &6.30&7.18&6.92 & 7.36\\
$^1{\rm A}_1$ &6.08 &5.92 &6.08 &6.08&7.02&6.72 & 7.41  \\
$^1{\rm B}_2$ &6.51 &6.36 &6.53 &6.52&7.37&7.12 & 7.49  \\
  & & & & & \\
m.e. &-0.63 &-0.79  &-0.62 &-0.61 &-0.23 &-0.39 &... \\
m.a.e. &0.70  &0.79 &0.64 &0.53 &0.24 &0.39 &... \\
\hline \hline
\end{tabular}

\footnotesize{$^a$From Ref.~\cite{gus99}.}
\end{table}
%%%%%%%%%%%%%%%%%%%%%%%%%%%%%%%%%%%%%%%%%%%%%%%%%%%%%%%%%%%%%%
\begin{table}
\caption{ Low-lying excitation energies (in eV) of ethylene
(C$_2$H$_4$)
calculated using six functionals with the basis set 6-311++G(3df,3pd).
Calculations are performed using the geometry optimized on respective
functionals with the same basis. The mean experimental value is 7.40 eV.}
\begin{tabular}{llccccccc}
\hline \hline
Symmetry & LSD & PBE & TPSS & TPSSh & PBE0& B3LYP & Expt$^a$
\\ \hline
$^3{\rm B}_{\rm 1u}$ &  4.81 &4.26 &4.12 &4.02 &3.97&4.17 & 4.36  \\
$^3{\rm B}_{\rm 3u}$ &  6.75 &6.45 &6.58 &6.74 &6.86&6.65 &6.98  \\
$^1{\rm B}_{\rm 3u}$ & 6.82  &6.58 &6.67 & 6.84&7.01&6.75 &7.15 \\
$^1{\rm B}_{\rm 1u}$ & 7.58  &7.44 &7.53 &7.59 &7.61&7.48 & 7.66  \\
$^3{\rm B}_{\rm 1g}$ & 6.95  &6.99 &7.17 &7.34 &7.39&7.27 & 7.79 \\
$^3{\rm B}_{\rm 2g}$ &  7.34 &7.02 &7.12 &7.31 &7.52&7.26 & 7.79\\
$^1{\rm B}_{\rm 1g}$ &  7.36 &7.16 &7.25 &7.43 &7.60&7.34 & 7.83\\
$^1{\rm B}_{\rm 2g}$ & 7.41  &7.13 &7.21 &7.40 &7.64&7.34 & 8.0\\
$^3{\rm A}_{\rm g}$  & 8.39  &8.03 &8.20 &8.33 &8.37&8.25 & 8.15\\
$^1{\rm A}_{\rm g}$  &8.71   &8.48 &8.56 &8.70 &8.85&8.63 & 8.29\\
  & & & & & \\
m.e. & -0.22 &-0.47 &-0.37&-0.25 &-0.12 &-0.29 &... \\
m.a.e.  & 0.41 &0.50 &0.42&0.35&0.27 &0.37 &... \\
\hline \hline
\end{tabular}

\footnotesize{$^a$From Ref.~\cite{pople92}.}
\end{table}
%%%%%%%%%%%%%%%%%%%%%%%%%%%%%%%%%%%%%%%%%%%%%%%%%%%%
\begin{table}
\caption{ Low-lying excitation energies (in eV) of benzene
(C$_6$H$_6$)
calculated using six functionals with the basis set 6-311++G(3df,3pd).
Calculations are performed using the geometry optimized on respective
functionals with the same basis. The mean experimental value is 5.89 eV.}
\begin{tabular}{llccccccc}
\hline \hline
Symmetry & LSD & PBE & TPSS & TPSSh & PBE0& B3LYP & Expt$^a$
\\ \hline
$^3{\rm B}_{\rm 1u}$ &4.47 &3.98  &3.84 &3.73 &3.68&3.84 &3.94  \\
$^3{\rm E}_{\rm 1u}$ &4.82 &4.61  &4.67 &4.70 &4.75&4.72 &4.76 \\
$^1{\rm B}_{\rm 2u}$ &5.33 &5.22  &5.32 &5.42 &5.52&5.41 &4.90  \\
$^3{\rm B}_{\rm 2u}$ &5.05 &4.89  &4.98 &5.06 &5.12&5.07 &5.60  \\
$^1{\rm B}_{\rm 1u}$ &6.07 &5.94  &6.00 &6.09 &6.18&6.05 &6.20  \\
$^1{\rm E}_{\rm 1g}$ &6.12 &5.89  &5.99 &6.18 &6.38&6.11 &6.33 \\
$^3{\rm E}_{\rm 1g}$ &6.09 &5.84  &5.95 &6.14 &6.32&6.07 &6.34 \\
$^1{\rm A}_{\rm 2u}$ &6.70 &6.43  &6.50 &6.69 &6.90&6.62 &6.93\\
$^1{\rm E}_{\rm 2u}$ &6.71 &6.44  &6.50 &6.70 &6.95&6.65 &6.95\\
$^3{\rm E}_{\rm 1u}$ &6.66 &6.37  &6.45 &6.63 &6.82&6.57 &6.98\\
  & & & & & \\
m.e. & -0.09 &-0.33 &-0.27&-0.16&-0.03 &-0.18 &... \\
m.a.e.  & 0.30 &0.40 &0.36&0.26&0.17 &0.28 &... \\
\hline \hline
\end{tabular}

\footnotesize{$^a$From Ref.~\cite{pdejo96}.}
\end{table}
%%%%%%%%%%%%%%%%%%%%%%%%%%%%%%%%%%%%%%%%%%%%%%%%%%%%%%%%%%%%%
%%%%%%%%%%%%%%%%%%%%%%%%%%%%%%%%%%%%%%%%%%%%%%%%%%%
\begin{table}
\caption{ Low-lying excitation energies (in eV) of pyridine
(C$_5$H$_5$N)
calculated using six functionals with the basis set 6-311++G(3df,3pd).
Calculations are performed using the geometry optimized on respective
functionals with the same basis. The mean experimental value is 5.07 eV.}
\begin{tabular}{llccccccc}
\hline \hline
Symmetry & LSD & PBE & TPSS & TPSSh & PBE0& B3LYP & Expt$^a$
\\ \hline
$^3{\rm B}_1$ &3.69 &3.68 &3.84 &3.99&3.81&3.97 & 4.1  \\
$^3{\rm A}_1$ &4.59 &4.11 &3.97 &3.86&4.08&4.05 & 4.1 \\
$^1{\rm B}_1$ &4.22 &4.33 &4.55 &4.74&4.86&4.76 & 4.59 \\
$^3{\rm B}_2$ &4.62 &4.41 &4.44 &4.49&4.54&4.52 & 4.84 \\
$^3{\rm A}_1$ &5.04 &4.78 &4.81 &4.86&4.92&4.88 & 4.84 \\
$^1{\rm B}_2$ &5.46 &5.33 &5.41 &5.53&5.63&5.52 & 4.99 \\
$^3{\rm A}_2$ &4.19 &4.30 &4.57 &4.83&5.03&4.93 & 5.40 \\
$^1{\rm A}_2$ &4.29 &4.43 &4.71 &4.99&5.20&5.07 & 5.43 \\
$^3{\rm B}_2$ &5.45 &5.40 &5.65 &6.06&5.72&5.64 & 6.02$^*$ \\
$^1{\rm A}_1$ &6.03 &5.97 &6.18 &6.31&6.41&6.23 & 6.38 \\
  & & & & & \\
m.e. &-0.31  &-0.40 &-0.26 &-0.08 &-0.05 &-0.11 &... \\
m.a.e.  &0.54  &0.47 &0.34 &0.25 &0.25 &0.26 &... \\
\hline \hline
\end{tabular}

\footnotesize{$^a$From Ref.~\cite{pople92}; $^*$CASPT2 estimate
from Ref.~\cite{lfr95,gus99}.}
\end{table}
%%%%%%%%%%%%%%%%%%%%%%%%%%%%%%%%%%%%%%%%%%%%%%%%%%%%
\begin{table}
\caption{ Mean absolute relative error (m.a.r.e.) of the atoms and molecules
listed in Tables I--IX. } 
\begin{tabular}{llcccccc}
\hline \hline
 &LSD & PBE & TPSS & TPSSh & PBE0& B3LYP 
\\ \hline
m.a.r.e. (\%) & 7.3 & 8.3 & 7.1 & 5.7 &4.4 &5.3 \\
\hline \hline
\end{tabular}
\end{table}
%%%%%%%%%%%%%%%%%%%%%%%%%%%%%%%%%%%%%%%%%%%%%%%%%%%%

Tables II--IV display the vertical excitation energies of three prototype 
inorganic molecules CO, N$_2$, and H$_2$O. For the CO molecule, 
as shown in Table II, the adiabatic TPSS functional produces
the vertical (low-lying) excitation energies in better agreement with the
experimental values~\cite{nielsen80} than the adiabatic 
PBE GGA, while it is slightly less accurate than the adiabatic LSD.  
As expected, the adiabatic TPSSh yields further improvement over the TPSS
meta-GGA. Partly mixing some amount of the exact exchange into a semilocal 
functional improves the asymptotic behavior of the XC potential. 
Similar results are observed for the N$_2$ molecule, an iso-electron series 
of the CO molecule. As observed in Table IV, both TPSS 
and TPSSh functionals describe the vertical excitations of water 
molecule well and produce the low-lying excitation energies more accurately 
than the adiabatic LSDA and PBE GGA. As expected, the best results are 
obtained with the adiabatic hybrid functionals PBE0, B3LYP, and TPSSh. 
We can see from the mean errors in Tables II--IV 
that all six adiabatic density functionals tend to underestimate the 
molecular excitation energies.

Tables V--IX show the low-lying excitation energies of five organic molecules
formaldehyde, acetone, ethylene, benzene, and pyridine. The adiabatic TPSS
meta-GGA consistently provides a more realistic description of the 
excitation energies of molecules than the adiabatic PBE GGA, and shows an 
overall improvement over the adiabatic LSDA. The adiabatic TPSSh gives  
further improvement upon the adiabatic TPSS functional, with  
comparable accuracy of the adiabatic PBE0 and B3LYP functionals.
Again as
we have already observed in Tables II--IV, these adiabatic density functionals
tend to underestimate the vertical excitation energies of molecules. 

In order to give an overall order of accuracy for these functionals tested here, 
we calculate the mean absolute relative errors of these functionals.
They are computed as follows. First we calculate the mean experimental value of 
the excitation energies of the atoms listed in Table I, as given in the caption. 
We also evaluate the mean experimental value of the excitation energies of each 
molecule, as shown in the caption of each table. Then we find the relative error
of each density functional from the ratio of the mean absolute error to the
mean experimental value. Finally we obtain the mean relative error by dividing
the sum of the nine mean absolute errors by nine, the number of the mean 
absolute errors of each functional. In short, the mean absolute relative error
is calculated as
\begin{eqnarray}
{\rm m.a.r.e} =
\frac{\Sigma[{\rm m.a.e.}/{\rm mean~ expt.~ value~ of~ property}]}{
{\rm number~of~the~ mean~absolute~ errors }}.
\end{eqnarray}
Table X shows the mean absolute relative errors of these functionals. We can see
from Table X that the overall order of accuracy for these functionals is
\begin{eqnarray}
{\rm PBE} < {\rm LSDA} \lesssim {\rm TPSS} < {\rm TPSSh} 
\lesssim {\rm B3LYP} < {\rm PBE0}.
%{\rm PBE} < {\rm LSDA},~ {\rm TPSS} < {\rm TPSSh},~ {\rm B3LYP} < {\rm PBE0}.
\end{eqnarray} 
The mean absolute relative error of each density functional tested here is less
than $10 \%$, suggesting the good performance of the adiabatic TPSS and TPSSh 
functionas for the description of low-lying excitations of atoms and molecules. 

The systematic underestimate of the low-lying vertical excitation energies
of molecules with the adiabatic approximation within TDDFT
suggests that further improvement can be made
with nonadiabatic corrections, as found by 
van Faassen et al.~\cite{Faa02,Faa04} with the Vignale-Kohn 
current-density functional theory~\cite{vk96,vuc}. The nonadiabatic
corrections for the inhomogeneous system have been recently 
derived by Tao and Vignale~\cite{tv06,tvt071}.    
%%%%%%%%%%%%%%%%%%%%%%%%%%%%%%%%%%%%%%%%%%%%%%%%%%%%

\section{Conclusion}
In this work we have tested the ability of the 
time-dependent density functionals TPSS meta-GGA and TPSSh hybrid 
within the adiabatic approximation to describe the low-lying 
excitations of atoms and small molecules. The results show that both
density functionals produce the vertical excitation energies
in fairly good agreement with experiments and improve upon the 
adiabatic LSDA and in particular, the adiabatic PBE GGA. This suggests
that both TPSS and TPSSh functionals within the simple adiabatic
approximation are capable of describing photochemically interesting
phenomena when the system is exposed to a time-dependent 
laser field.
Compared to other adiabatic density functionals, the adiabatic TPSS
functional yields the best performance among nonhybrid functionals,
while the adiabatic TPSSh functional can achieve the comparable accuracy
of the most popular hybrid functionals B3LYP and PBE0. Further 
comprehensive tests for larger molecular 
systems~\cite{gus99,sergei1,gus071,yang07} are necessary in
order to assess performance of these functionals that have been developed
for ground-state properties, but not for TDDFT applications to 
excited states, for new photophysical and photochemical phenomena
(such as charge transfer present in nanosystems).

In view of the good performance of the TPSS functional for diverse 
systems and a wide class of properties, we conclude that TPSS 
is indeed a reliable nonhybrid universal functional,  which can serve 
as a platform from which higher-level approximations can be constructed.
The advantage is that we are able to use the same method and the
same basis set to simultantiously describe different class of problems 
such as chemical reactions on metal surfaces.

\acknowledgments
%\section*{ACKNOWLEDGMENTS}
We thank Ping Yang,  Ekaterina Badaeva, and Xinzheng Yang for technical help, 
and Richard Martin, Gustavo Scuseria, and John P. Perdew for useful 
discussions.  This work was carried out under the auspices of the National
Nuclear Security Administration of the U.S. Department of
Energy at Los Alamos National Laboratory under Contract No.
DE-AC52-06NA25396 and under the Grant No. LDRD-PRD X9KU.

%%%%%%%%%%%%%%%%%%%%%%%%%%%%%%%%%%%%%%%%%%%%%%
\end{document}